# A system and methodologies for absolute QE measurements from the vacuum ultraviolet through the NIR

Blake C. Jacquot,[1] Steve P. Monacos,[1] Michael E. Hoenk,[1] Frank Greer,[1] Todd J. Jones,[1] Shouleh Nikzad[1]

[1]Jet Propulsion Laboratory, California Institute of Technology, 4800 Oak Grove Drive, Pasadena, CA, USA 91109

## ABSTRACT

In this paper we present our system design and methodology for making absolute quantum efficiency (QE) measurements through the vacuum ultraviolet (VUV) and verify the system with delta-doped silicon CCDs. Delta-doped detectors provide an excellent platform to validate measurements through the VUV due to their enhanced UV response. The requirements for measuring QE through the VUV are more strenuous than measurements in the near UV and necessitate, among other things, the use of a vacuum monochromator, good dewar chamber vacuum to prevent on-chip condensation, and more stringent handling requirements.

## I. INTRODUCTION

Accurate measurement of absolute quantum efficiency (QE) requires careful attention to a variety of system design parameters and measurement methods. Compounding the difficulty in making absolute QE measurements is the added requirement of testing through the vacuum ultraviolet (VUV, corresponding to the 100-200 nm spectral range). This range introduces a number of special considerations and methodologies that add materially to the requirements for the near ultraviolet and visible wavelengths.

For our QE measurement test apparatus, we use a custom vacuum monochromator equipped with a dual tungsten and deuterium source that is selected by an index mirror, reference and calibrated diodes, a filter wheel for further spectral selection, and diffusers to provide flat-field illumination to either a CCD or a calibrated photodiodes (Fig. 1). These and other components and our procedures to use them are described throughput this paper. Immediately before and after the CCD or calibrated photodiode measurement a reading is taken from a reference silicon photodiode selected by a diverter mirror to mitigate any time-based source fluctuations. The ratio of this photodiode with the reading from the calibrated photodiode set provides the information needed to interpret results from the CCD. In this work, we present the results of testing delta-doped CDDs in this system and show that our measurement and calibration methods give the expected reflection-limited response throughout far UV to near infrared (NIR). by using the above method. One of the most challenging aspects of absolute QE measurement is an accurate determination of photon flux. For that we rely on the NIST and IRD-calibrated photodiodes.

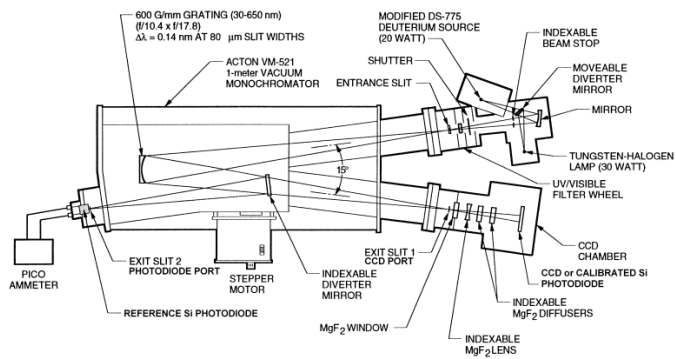

FIG. 1. UV/Visible QE test apparatus

This global calibration approach negates the need to test individual components of the light path, as is sometimes done. The flat field allows all pixels to see approximately the same illumination and also allows calibration of multiple CCD sections simultaneously. An alternative approach may use a spot illumination and would require the movement of the CCD to measure QE in all sections. An integrating sphere can also provide a flat field, but common materials for coating the interior absorb in the VUV. Reportedly, a VUV compatible integrating sphere is in use at the European Southern Observatory (ESO) and at JPL has been used for UV instrument calibrations such as GALEX mission.

The reference diode is measured by an electrometer (Keithley 6517A). It is taken in and out of the light path by a hand-driven diverter mirror. Tuning to a given wavelength is performed using a computer-controlled stepper grating adjustment and dialing in the correct filter. We have observed that once the source lamp is warmed up for an hour, the lamp drift has negligible impact on the measurement. This is due to the independent calibration set described above which compares the light at both ports in near real time. The tests are bracketed in time by a before and after reading from the reference diode to check for signal drift. These two reading are averaged together to represent flux seen by the detector. The signal typically drifts only by less than 0.5%. Thus, we do not correct for lamp drift with a feedback loop involving an uncalibrated photodiode at the source and variable voltage control.

In this paper we detail our approach at JPL for conducting QE measurements from the VUV through the visible and NIR. In section II systematic error mitigation is discussed. Section III discusses system calibration. Section IV discusses camera and environmental considerations. Section V presents signal interpretation and detector electronics. Section VI discusses QE determination and quantum yield. Section VII talks about delta doping and device preparation. Finally, Section VIII presents results.

## II.     MITIGATION OF SYSTEMATIC ERROR

Temporal averaging can suppress statistical errors and so systematic errors, discussed below, limit ultimate accuracy of a measurement. For example, even a small amount of out-of-band light leak can perturb a measurement significantly. For example, if a filter allowed 100% transmission at a UV wavelength (e.g. 350nm) and only 0.1% transmission for a visible light wavelength (e.g. 650nm), the resulting QE error could be substantial when using a Tungsten-Halogen source. The source emits approximately 100 times more photons in the visible than the UV. Typical QE for reference detectors is approximately 40% in the UV and 90% in the visible. This would give a signal of 40 electrons resulting from UV light for every 9 electrons resulting from visible light. This comes from a 1000x reduction of visible relative to UV due to light rejection from filtering but a 100x brighter region of the source and 90% conversion gain between photons and electrons. In this example, the electrons generated by a miniscule red leak equal nearly 20% of total signal. Systematic errors (including especially out-of-band illumination) can be controlled by source selection, filtering, reference detector selection, careful design of the optics and optical path, using diffusers for field flattening, and careful system calibration.

### A.  Sources

Proper choice of light sources aids in the elimination of out-of-band leak. Red leak, the most commonly-discussed type, results from any wavelength longer than the desired range contaminating the signal. Causes of red leak include poorly filtered light from a source with much stronger emission at wavelengths longer than the measurement wavelength, unwanted reflections and scattering (i.e. including imperfections and contamination on the grating and/or mirrors), imperfect baffling of off-axis light, glow from components such as ion gauges, and wavelength ranges where order-sorting filters do not provide adequate discrimination. Blue leak can also be troubling if unmanaged. Contamination from wavelengths shorter than the desired wavelength (e.g. from strong UV sources) or second order light from a monochromator can introduce error.

To guard against out-of-band leak, we have chosen to use a deuterium source for illumination from Lyman-alpha to approximately 350nm and a quartz tungsten halogen source from 360nm to the near infrared (Fig 2). A hand-driven flip mirror selects the source and directs the light through a single filter set (Fig. 1). The deuterium source has the advantage of strong emission between 100-200nm, a lesser, relatively flat emission profile to approximately 450nm, and, importantly for minimizing red leak, virtually no emission intensity in the visible and near infrared regions of the spectrum. The quartz tungsten halogen source exhibits a blackbody radiation profile with a peak around 800-1000 nm, and very low emission intensity in the ultraviolet, thus minimizing blue leak.

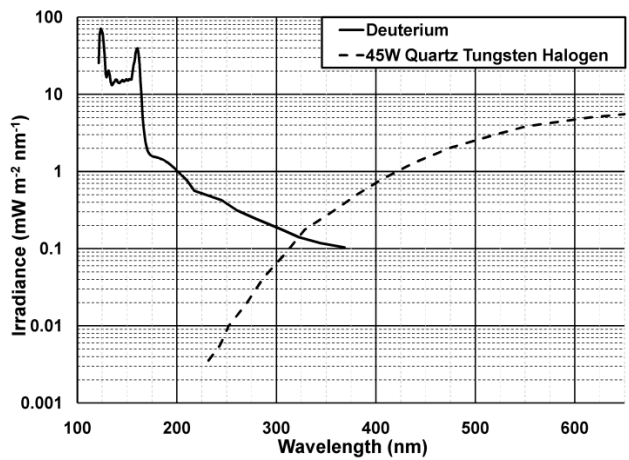

FIG. 2. Representative irradiance of sources. Note wavelength range between 250nm and 400nm where neither source is strong. Use of quartz tungsten halogen source in this range is highly susceptible to red leak if aggressive bandpass filters are not employed.

In regions where the sources are strong (e.g. below 200nm for Deuterium and 500-900 nm for Tungsten-Halogen) guarding against out-of-band leak becomes simpler due to the high signal relative to out-of-band wavelengths. However in regions where neither source has strong emission (e.g. 250nm-400nm) one must carefully choose filters to ensure light leak does not occur or at least is not an appreciable fraction of the total signal.

## B. Filtering Overview

Beyond source selection, the choice of filters is the other major consideration in the elimination of systematic error. As mentioned above, out-of-band leak can destroy a high-quality measurement. Some origins of out-of-band leak include stray light reflections off walls of the monochromator, stray light reflection off the monochromator grating, and second-order light contamination from the grating. Properly chosen bandpass and longpass filters can dramatically reduce most of these sources of error.

The first concern when choosing a filter is consideration of second-order light from the monochromator. Longpass filters for use with monochromators (commonly called "order sorting" filters) provide excellent protection from this phenomenon. Monochromator gratings allow both the first order light containing the wavelength of interest and the second order containing half the wavelength of interest to reach the detector.[1] For instance a desired wavelength of 500nm will also contain a smaller amount of 250nm light. Filters should be chosen to at least eliminate the second order light. Coming after these filters, the monochromator grating effectively acts as a tunable band-pass filter which selects a narrow band of light for illumination of the detector or photodiode. The combination of front-end filtering and the filtering provided by the grating effectively limits out-of-band light leak.

Our measurement setup (Fig. 1) allows for a 7-position filter wheel which we populated with 6 filters and an open position, as shown in Table 1 with the monochromator grating acting effectively as an additional tunable narrowband bandpass filter. For the deuterium source, uncoated Fused Silica and Pyrex serve as long-pass filters (obtained from Princeton Instruments), or filterless openings. The deuterium spectrum below 164nm is relatively strong, relieving concerns about red leak. Fused Silica and Pyrex are longpass filters that guard primarily against blue leak from the deuterium source. Note that Fused Silica, with its approximately 160nm cutoff, can be used up to twice this wavelength before second order light starts to contaminate the signal, so we discontinue its use around 300nm. The same reasoning applies to Pyrex, though twice its cutoff of approximately 300nm is well beyond the usable range of the deuterium source. We employ bandpass filters in the 250-400nm region where neither source is strong. The relatively narrowband U-340 filter with 85nm passband provides excellent discrimination against red leak from the Tungsten-Halogen source. The B-390 filter has a larger passband of 125nm and also provides filtering against red leak. The longpass GG-420 filter provides order sorting filtering for the Tungsten-Halogen source up to approximately 840nm (double the cutoff). However the grating used in this measurement does not extend beyond 650nm. We also include an RG-630 longpass filter, but it is not used in current data collection. The U-340, B-390, GG-420, and RG-630 filters were obtained from Edmund Optics.

TABLE 1. Filters and sources used in QE characterization

| Filter | Utilized Range (nm) | Note | Source |
|--------|---------------------|------|--------|
| None | 116-164 | Must limit exposure due to high energy of radiation. Out of band leak <0.5% of total signal. | Deuterium |
| Fused Silica | 164-300 | Long-pass filter with cutoff at ~160nm. Out of band leak <0.5% of total signal. | Deuterium |
| Pyrex | 300-350 | Long-pass filter with cutoff at ~300nm. Out of band leak <0.5% of total signal. | Deuterium |
| U-340 | 360-370 | Hoya absorptive bandpass filter, 85nm passband with 340nm center wavelength. Red leak accounts for ~2% of total signal. | Tungsten-Halogen |
| B-390 | 380-450 | Hoya absorptive bandpass filter, 125nm passband with 390nm center wavelength. Out of band leak <1.4% of total signal. | Tungsten-Halogen |
| GG-420 | 460-650 | Schott absorptive longpass filter with cutoff at ~420nm and full transmission starting at ~495nm. Out of band leak <10% of total signal. | Tungsten-Halogen |
| RG-630 | 660-1100 | Schott absorptive longpass filter with cutoff at ~630nm and full transmission starting at ~690nm. Out of band leak not tested. | Tungsten-Halogen |

Efforts to measure out-of-band leak indicate the above filtering scheme is successful. Experimenters may use the following methodology to attempt quantification of out-of-band leak. The Tungsten-Halogen source allows for convenient characterization since a separate filter can be placed in front of the source to limit light reaching the filter wheel. For example, when tuned to 340nm with the U-340 filter selected in the filter wheel and a separate GG-420 2" square longpass filter sitting in front of the source, the total current read by the a picoammeter was 0.1pA as compared to 4.8pA when the GG-420 filter was removed. This indicates light leak above the 420nm cutoff of the longpass filter contributes approximately 2% to the total signal. Similar tests for the B-390 filter (also using the GG-420 filter) when tuned to 390nm indicate an out-of-band leak accounts for <1.4% of total signal. Tuning to 450nm and using the GG-420 filter in the 7-position filter wheel with a separate longpass filter with 630nm cutoff indicates out-of-band leak accounts for <10% of total signal, an unacceptable amount. The signal to noise ratio improves dramatically when moving to longer wavelengths due to increasing strength of the source and is the primary reason why 500nm was selected as the transition point between the B-390 and GG-420 filter. Out-of-band leak tests for the Deuterium source comparing readings from using a Fused Silica and Pyrex filter at 121.6nm, 200nm, and 250nm propitiously indicate out-of-band leak contributes less than 0.5% of total signal.

Broadband reflection off gratings or other components in the optical path will also result in signal contamination and requires filtering. Directly observing the grating when tuned to 500nm will show a bright green light with specks of white light contamination. Since we illuminate either a reference diode or CCD with the same light path, the best approach for our system involves using bandpass filters at the source to restrict the range of light allowed into the monochromator. The more narrow the bandpass filter, the greater the purity of light at the desired wavelength. However, practical considerations of the available physical space for filters (i.e. filter wheel positions) may necessitate acceptable tradeoffs between filter choice and optimization.

## C.  Reference Detectors

The reference detector and calibrated silicon photodiodes in our setup are monitored with a Keithley 6517A electrometer. The noise and dark current error are negligible with respect to the measurement. Typical values as read by the electrometer have magnitudes of less than 0.01 pA under no illumination. This compares well with readings in the 10s to 1000s of pA when illuminated. As such we do not employ a chopper to reduce thermal effects. The diodes are operated at room temperature with no active cooling. The electrometer is warmed up for at least an hour prior to measurement to ensure constant temperature.

We have seen that even very small red leak can lead to significant out-of-band systematic error. The choice of reference detectors can help reduce measurement error. It is preferable that the device under test and the reference detector have approximately the same spectral response so that red leak will lead to equivalent error in both and thus divide out. Often CsTe photocathodes are used since they are blind to visible light. However, CsTe may not be optimal choice because it has differing response relative to silicon towards out-of-band illumination. If the spectral efficiency of the device under test is identical to that of the reference detector, the systematic errors due to out-of-band illumination of the detector may be offset by equivalent response of the reference detector, and there will be relatively small systematic errors due to out-of-band illumination. In this case, both measurements will include an identical offset from the red (or blue) leak. However if the spectral responses of detector and reference detector differ greatly (as with a CsTe reference), the resulting systematic error can be unexpectedly large.  From this reasoning, one may make the argument that the detector under test and the chosen calibrated standard should be made of the same material (e.g. silicon) to attempt a relatively close matching of spectral responses. As opposed to silicon, the visible blind nature of CsTe photocathode may result in error of measured QE for a silicon CCD, since the CCD has broadband response.

We have chosen broadband silicon photodiodes as our calibrated standards due to their large wavelength range and the desire to not change diodes during measurement, thus reducing opportunities for hydrocarbon contamination and human error. During initial system calibration two photodiodes are used in place of the CCD imager to measure light levels as seen by the CCD. A NIST-calibrated IRD AXUV-100G is used to measure light between 116-254nm. A second IRD-calibrated UVG-100 photodiode is used to measure between 250-1100nm. A separate uncalibrated IRD AXUV-100G sits permanently at another port within the monochromator and provides a reference reading for system calibration. In principle the IRD AXUV-100G can cover the entire wavelength range desired, but NIST calibration lead time constraints necessitated the 2-diode approach.

As with other components, the photodiode response is very sensitive to hydrocarbon contamination. Further, the passivating oxides can be damaged by high-energy UV radiation.[2-3] We limit UV exposure to avoid damage and periodically re-calibrate our standards and so we have confidence in the flux measurements with NIST-calibrated diodes. NIST typically screens photodiodes from lot run to determine expected damage from UV exposure.

## D.  Optics and Optical Path Components

Careful selection of optics and components is needed for compatibility with VUV illumination. $MgF_2$ is the most common material for transmissive windows and coatings and is used extensively in our system. Mirrors use aluminum coated with $MgF_2$ as a protective coating. Many of the materials will degrade with prolonged exposure to UV light, necessitating periodic recalibration. UV light exposure should be minimized when possible to avoid this damage. Due to the rigorous filtering scheme adopted and the selectivity of the grating, we do not expect decrease in S/N or an increase in systematic error. Periodic re-calibrations show the system to be exceedingly stable with respect to wavelength over time.

Reflection from sidewalls can cause error in measurement and so baffles should be deployed to collimate the light wherever possible. This becomes more crucial after the diffusers and before the detector due to the spherical emanation of light which can easily reach and reflect from sidewalls. We use a series of 3 baffles spaced 1" apart prior to the CCD to collimate the light.

## E.  Flat Field Illumination

In many of our measurements we want to compare various antireflective (AR) coatings and bare surfaces on the same imaging device, demanding a flat field so that all surfaces see equal illumination. A flat field is also a good way to evaluate uniformity of the device itself. We realize a flat field by using two $MgF_2$ diffusers of 1" diameter which sit at the end of the monochromator signal chain and approximately 6" in front of the detector. This configuration allows for less than 5% light intensity variation across an approximately 2cm x 2cm surface, as measured directly by a CCD. This profile is consistent across a variety CCD models, leading to the assumption it stems from the illumination path rather than variable pixel response. The light intensity variation is closer to 1% in the central 1cm x 1cm region used for CCD calibration and calibrated photodiode measurement, allowing for high quality QE measurement. Often an integrating sphere is used to create a flat field, but this is difficult for VUV measurements due to material constraints and need for a vacuum environment. Variations of polytetrafluoroethylene (PTFE, e.g. Teflon) are commonly used for coating the inside of spheres and providing a reflective surface. While PTFE works well for visible light, it absorbs in the VUV. Adding further complication, most integrating spheres are not vacuum compatible. Though we had discussions with a vendor about making a vacuum-compatible integrating sphere with Al coating, the costs were prohibitive and we were able to obtain excellent flat fields with the above methods.

# III. SYSTEM CALIBRATION

## A. Throughput Calibration

We rely on a global method of calibrating the measured light rather than a component-by-component method. Our procedure consists of tuning to a given wavelength and measuring the reading from a calibrated photodiode sitting in place of the target CCD and comparing this reading to a reference photodiode at a separate port selected by a diverter mirror in the monochromator. This near real-time measurement provides a ratio of the light at the two ports which can then be used to interpret readings from the CCD. For instance, if at 200nm we see a reading of 10pA from the CCD port diode and 175pA at the diverter mirror selected port, the ratio (10pA/175pA) can be used to determine the light reaching the CCD when we have a fresh reading from the diverter mirror selected port during the QE measurement. The profile of the diffused light at the CCD port and profile of the non-diffuse light at the photodiode port remain constant, and can be characterized to avoid systematic error.

The purpose of the reference photodiode selected by the diverter mirror is to measure and correct for lamp intensity (or upstream throughput) variations between measurements made with the CCD and those made with a calibrated photodiode in place of the CCD. The characterization between ports is accomplished by adopting a consistent diffuser element and slit size configuration at the CCD port, measuring the photon flux at this port using a calibrated photodiode, and measuring the aggregate photon signal for the non-diffuse light with the reference photodiode. The photon flux incident on a CCD pixel element is determined from a reference photodiode reading. This reading is converted to the commensurate photon flux at the CCD port and then scaled from the photodiode area to the pixel size to determine the photons incident on a pixel element. Since the reference photodiode reading acts as a transfer standard only, it does not introduce a systematic error into the results. Using this method, we typically see that sources exhibit less than 10% variation in light level from measurement to measurement. This variation does not affect QE measurement due to the inherent reduction of 1/f noise from this measurement scheme. Both the calibrated photodiode and CCD alternately sit at the same position for calibration. QE measurements have shown the system to be very tolerant to variations in physical position from one to the other. For instance, a 0.25" position variation between the two gives negligible QE error. We illuminate approximately a 1mm x 1mm portion of the reference photodiode as compared to 100% of the 1cm x 1cm of the same model NIST-calibrated photodiode at the CCD port. The fill is not important for the reference photodiode, nor is its long-term stability. Given frequent re-calibrations the QE drift of the reference photodiode will drop out of the measurement since it is used as a near real-time transfer standard only. In any case, we see only negligible system drift over the course of 12 months of heavy use. The other prime source of system drift will come from the change in reflectivity of surfaces, such as the grating or diverter mirror. As mentioned, we have not observed this to occur in an appreciable amount.

Another common method of system calibration is to measure the transmission and reflectance of each element along the signal chain and compile them for a global throughput. We have performed this style of single component calibration and it also can be periodically re-calibrated. However, this method does not work effectively if the final measurement requires a flat field. The flat field (either created by diffusers or an integrating sphere) is most easily and accurately measured directly.

## B. Wavelength Calibration

Another step in system calibration is determining the tuning of the monochromator grating wavelength against programmed mechanical counter reading. To accomplish this, several known wavelengths must be determined and fit using a polynomial function. In our system we used Lyman-alpha (121.6nm) and several narrowband interference filters (340nm, 500nm, 640nm) for a total of 4 data points. We then used a 3rd order polynomial to fit the mechanical counter to wavelength data and interpreted the tuning of the monochromator with the results of this fit.

## IV. CAMERA AND ENVIRONMENTAL CONSIDERATIONS

### A. System Overview

Measurements from 100nm to 200nm need a vacuum monochromator for good data integrity due to atmospheric absorption.[4] Alternative approaches may use a nitrogen purged environment. At JPL, an Acton 1m vacuum monochromator with either 300grooves/mm or 600grooves/mm grating is used to characterize devices from Lyman-α (121.6nm) through the near infrared.

Beyond a vacuum environment, VUV measurements have great sensitivity to surface contamination.[5] Hydrocarbons in the vacuum system (e.g., pump oils) can condense on components of the system (e.g. grating, mirrors, windows, diffusers, diodes, etc.) and become permanently burned in place with VUV radiation. If not properly guarded against, the system state can shift and render calibration meaningless. Periodic re-calibration is necessary to monitor for this effect.

### B. Cleaning and Bakeout of Components

Any component entering the vacuum system should experience a substantial cleaning and bakeout procedure (aside from very sensitive components such as mirrors, gratings, and windows). Typically, we sonicate metals for 30 minutes with an ultrasonic cleaning solution (e.g. Fisherbrand or Alconox) followed by 5 minutes of acetone and 5 minutes of IPA sonication and finally a vacuum bakeout. Typical vacuum bakeout conditions are 120ºC (or higher) for 16-48 hours. Any organic entering the system (e.g. O-rings, printed circuit boards, electronic cables) should experience a similar cleaning procedure with the exception of the acetone bath. Bakeout of organics is extremely critical for minimizing outgassing in the vacuum system, which can irrevocably contaminate surfaces with hydrocarbons.

### C. Temperature and Vapor Pressure of Ice Effects and Other Condensates on Measured QE

Cryogenic measurements of CCDs require attention to the vacuum condition in order to avoid buildup of condensation or ice on the detector surface that degrades measurements. High-quality scientific measurements of CCDs require cooling the detector to minimize dark current, typically in the range of -130ºC to -80ºC. However as temperature drops, so does the vapor pressure of ice.[6-7] For instance, at -50ºC the vapor pressure of ice is approximately $3x10^{-2}$ Torr. If the camera base pressure is above this level, ice will deposit on the detector surface. At -80ºC the vapor pressure of ice is $4.1x10^{-4}$ Torr. Our camera base pressure currently operates at approximately $5x10^{-6}$ Torr or less, which roughly equals the vapor pressure of ice at approximately -90ºC. To allow for a margin of safety, we often choose an operating temperature of -80ºC to avoid ice buildup. Attempting to operate significantly colder (i.e. -130ºC) leads to systematic errors as illustrated by the data in Figure 3. To mitigate this problem, we are currently installing a dewar/camera designed to achieve base pressures in the $10^{-9}$ Torr range, which will enable "ice-free" characterization of CCDs down to -130ºC or below. Investigating the data, we saw that Trial 1 showed spurious behavior with respect to temperature, so we performed a thermal cycle and repeated the test in a more controlled fashion in Trial 2. Comparing Trial 1 measurements (made from high wavelength to low) and Trial 2 measurements (low to high wavelength) shows hysteresis both in VUV and visible regions, with VUV most pronounced. The hysteresis affects all wavelengths, but shows time dependence as absorbing ice builds up on the detector surface. Measurements made at approximately 220 nm had seen cryogenic temperatures for roughly the same amount of time, accounting for their similar values.

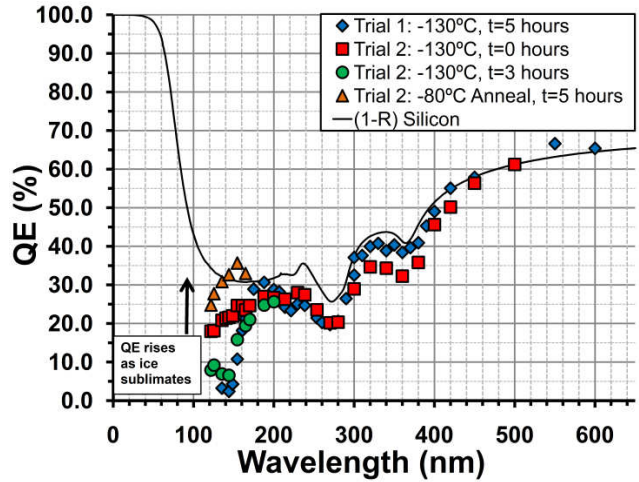

FIG. 3. Demonstration of measurement hysteresis when operating at -130ºC with ice as the likely cause. Trial 1 showed spurious behavior with respect to temperature, so we performed a thermal cycle and repeated the test in a more controlled fashion in Trial 2. Measurements made after holding temperature at -130ºC for extended period of time are depressed relative to initial measurements at that temperature or after -80ºC anneal, which results in ice sublimation. The ice accumulation resulted from the camera base pressure exceeding the vapor pressure of ice at this -130ºC. Comparing Day 1 measurements (made from high wavelength to low) and Day 2 measurements (low to high) shows hysteresis both in VUV and visible regions, with VUV most pronounced. The hysteresis affects all wavelengths, but shows time dependence as absorbing ice builds up on the detector surface. Measurements made at approximately 220 nm had seen cryogenic temperatures for roughly the same amount of time, accounting for their similar values. Base pressure of camera relative to operating temperature must be considered to avoid this effect.

As demonstrated above, UV measurements are extremely sensitive to ice accumulation (or any other undesired surface layers) due to very short absorption depths of the photons. Extrapolating from formulas governing water vapor pressure, we calculate that at -130ºC the vapor pressure of ice is about $6x10^{-9}$ Torr. In other words, to operate at -130ºC one would need to achieve a camera base pressure in the mid to low $10^{-9}$ Torr range to ensure ice does not interfere with the high precision measurement of UV QE. However, note that at these exceptionally low pressures, even if ice condensation is occurring it will happen at a very slow rate. To achieve such low pressures ultrahigh vacuum (UHV) conditions should be met. This includes replacing O-rings with UHV flanges and discontinuing the use of all organics such as insulated wires.

At a minimum, efforts should be made to reach as low a base pressure as possible, which may include the use of getters, and to calculate the expected vapor pressure of ice for the temperature of operation. Exposure to atmosphere should be minimized to reduce re-absorption of water. Stainless steel should be used when possible due to its tendency to not absorb water. However, aluminum may be chosen for ease of machining as in the case of our dewar shell.

### D. Cryogenic Design

Our liquid nitrogen-based cryogenic design keeps the detector stable to ±0.2ºC. A copper cold finger assembly bolts to a roughly 1L liquid nitrogen tank within the interior of the dewar. A copper assembly connects the tank to the CCD. A proportional-integrative-derivative (PID) controller takes readings from a thermocouple located near the device and uses the information to drive a heater cartridge located in the copper assembly to stabilize the CCD temperature. Alternative approaches to this design may use a closed-cycle cooler (e.g. CRYOTIGER)

## V.  SIGNAL INTERPRETATION AND DETECTOR ELECTRONICS

### A.  Photon Transfer and System Gain

Photon transfer technique was developed by Jim Janesick as a method for measuring the system gain of CCD imaging detectors.[8] Accurate determination of system gain is required in computing QE. A photodiode readout converts easily to e-/sec while a CCD readout gives DN/sec so knowledge of the system gain (e-/DN) is needed to interpret results. The method is based on an extrapolation of detector shot noise to infer the system gain.  It rests on fitting a shot-noise limited curve and finding the x-intercept where the noise is unity in Digital Number (DN).[9] At this point, the x-intercept (or signal) is equal to the system gain. This extrapolation can result in substantial systematic error if the photon transfer curve is poorly constructed. A small change in the fit conditions can result in a relatively large change in perceived system gain. Noise sources must be treated properly when extracting the shot-noise limited curve. Note that the region of interest used for calculation of statistics must be large enough (e.g. 100 x 100) to beat down statistical errors which may propagate into QE. The references above provide a detailed discussion of the photon transfer procedure and nuances.

## B. Electronics

Electronic interface with the CCD is provided by an Astronomical Research Camera Inc. controller (i.e. "Leach Controller") which captures frame information from the detector.[10] A controller box is populated with a fiber optic timing board, a clock driver board, a utility board, and video boards. It communicates through a fiber optic cable with a PCI card residing in a personal computer. We collect FITs image files and import them into Matlab for data manipulation and processing to gather information for photon transfer and QE curves.

## C. Quantum Efficiency Determination

QE data were collected at JPL for a variety of delta doped devices. After generating a photon transfer curve for each device, flat fields were collected at a variety of wavelengths from vacuum UV through 600nm. For calculation of QE, a consistent 2D ROI was picked from the image and differenced from a similarly-timed dark integration. This ROI should correspond closely or exactly with that used for photon transfer in order to ensure the highest integrity of the resulting QE. The mean of the net signal in coordination with other factors such as the ratio of light between photodiodes, pixel size, photodiode reading, exposure time, photodiode size, and quantum yield are used to create a quantum efficiency plot versus wavelength

QE is found by relating the reading from a calibrated photodiode to the CCD response by

$$QE = \frac{A_D S_{CCD} QE_D}{A_P S_D} \quad (1)$$

where $A_D$ is the area of the calibrated photodiode (cm$^2$), $A_P$ is the active area of a pixel (cm$^2$), $S_{CCD}$ is the signal generated by the CCD (e-/pixel/sec), $S_D$ is the signal from the photodiode (e-/sec/cm$^2$), and $QE_D$ is the quantum efficiency of the photodiode. The experimenter should include appropriate scale factors to the above equation when appropriate to account for differences in illumination of the photodiode and device under test. The raw QE data should be adjusted for quantum yield below approximately 340nm.

# VI.   QE DETERMINATION AND QUANTUM YIELD

## A. Quantum Yield

QE is conventionally defined in terms of the ratio of collected electrons to the number of incident photons. For low-energy photons, there is no ambiguity or difficulty with this definition, as each detected photon generates a single electron-hole pair in the detector. At higher energies, incident photons carry enough energy to produce multiple electron-hole pairs, and the standard definition results in an apparent contradiction in which the measured QE can be significantly greater than expected. The most useful representation of detector efficiency accounts and normalizes for multiple electron-hole pair production in the detector. To accurately infer this measure of QE, a correction for quantum yield (QY) must be applied to the number of free carriers collected (Fig. 5).[11]An accurate determination of quantum yield for a detector is essential to correctly interpret experimental data in the far ultraviolet (FUV). It is a complex function of material, method of illumination, surface properties, and photon energy.

A common approximation for QY that is valid for high-energy photons involves making a linear approximation based on measured data from above 10eV.[9]

$$\eta_i = \frac{12390}{E_{e-h}\lambda} \quad (2)$$

Here, $\eta_i$ is the quantum yield gain (e-/photon), $E_{e\text{-}h}$ is the energy required to generate an electron-hole pair (approximately 3.65eV at room temperature for silicon), and $\lambda$ is the photon wavelength in Å.

Quantum yield has been directly determined in a variety of ways with varying results. Measurements from silicon photodiodes determine quantum yield by measuring reflectance loss and comparing photodiode readings to measured external quantum efficiency.[12-13] These measurements assume zero internal loss due to either absorption by the overlying silicon dioxide or recombination in the diode itself. Both assumptions are good down to approximately 150nm where the energy of photons roughly equals the bandgap of silicon dioxide and the material becomes absorbing, thus complicating the measurement. We have found the data from *Canfield, et. al.* provides a sensible interpretation of our raw QE values.[12] This digitized QY data fitted with a 6[th] order polynomial provides a good agreement between measured QE and the reflectance limit of silicon down to about 150nm. Below 150nm, the relationship breaks down and does not provide good translation.

The approximation used in this work is given as

$$QY = 1.26329 \times 10^{-12}\lambda^6 - 1.86609 \times 10^{-9}\lambda^5 + 1.14114 \times 10^{-6}\lambda^4 - 3.7039 \times 10^{-4}\lambda^3 \qquad (3)$$
$$+ 6.747379 \times 10^{-2}\lambda^2 - 6.5698494\lambda + 269.9114$$

Here $\lambda$ is wavelength in nanometers. The approximation works very well down to at least 150nm. An alternative way to obtain quantum yield is to measure it directly.[14] In this method the shot-noise limited intercept (i.e. std(DN)=1) is found for a wavelength above 340nm (e.g. 400nm) and ratioed with the intercept for a wavelength where quantum yield is greater than unity. A full photon transfer curve must be constructed for each wavelength to accurately measure the quantum yield, dramatically increasing the amount of data for an experiment and also risking UV damage of the detector. We have also used this technique and details will be reported separately.

## VII.    DELTA DOPING AND DEVICE PREPARATION

### A.  Delta Doping

Due to their high QE, enhanced UV response, and stability over a wide spectral range (EUV-NIR), delta-doped detectors provide an excellent platform for testing and validating QE data through the VUV. Unlike most devices, the delta-doped CCD UV and VUV response is predictable and reflection-limited. If one measures a silicon reflection-limited response from a delta-doped CCD the result is an indication not only of great device performance but also a validation of the measurement system and not simply an optimistic measurement making a fair device look better. Delta-doped CCDs, developed at JPL's Microdevices Laboratory, have achieved stable 100% internal quantum efficiency in the visible, near UV, and vacuum UV regions of the spectrum.[15-21] In this approach, an epitaxial silicon layer is grown on the back surface of a fully-fabricated CCD or CMOS back-illuminated imager using molecular beam epitaxy (MBE).[19] During the growth, approximately a third of a monolayer of dopant atoms is deposited on the surface, followed by growth of a silicon cap layer. The dopant is incorporated within a single atomic layer near the back surface of the device, resulting in the effective elimination of the backside potential well. The measured quantum efficiency is in good agreement with the theoretical limit imposed by reflection from the Si surface. Backside treatment is essential to keep free carriers away from energy traps at an unpassivated back surface. Further details can be found in the above references.

### B.  Backside Illumination

Front-illuminated devices present challenges for high quantum efficiency, thus necessitating the use of backside illumination for the most demanding applications. Front illumination is generally defined as illuminating the metalized side of a detector. Because of absorption and scattering of photons in the metals and gate oxides (leading to less than 100% fill factor), front-illuminated CCDs have limited QE and spectral range. Further, CMOS imagers have part of the pixel devoted to signal-processing circuitry, and the maximum achievable quantum efficiency is additionally limited by the fraction of the detector surface that is sensitive to light (fill-factor). Backside illuminated devices can dramatically improve QE by illuminating the non-metalized side of a detector. Back-illuminated imagers require thinning or high resistivity substrates with full depletion. A surface treatment such as delta-doping provides passivation as well as a backside electrode for full depletion in high-resistivity devices.

### C.  Thinning

Backside thinning is required to remove substrate material and prepare the device for processing enabling backside illumination. The removal of material allows the photosensitive region to approach or reach full depletion. Without this the QE may suffer due to carrier recombination and the device resolution will degrade due to carrier diffusion in a field-free region. Fully-processed CCD die, wafer, or rafts comprised of many contiguous die, complete with aluminum contacts, are thinned by either wet or dry etching, as displayed in Figure 4. In one wet etching, some 90% of the bulk silicon is removed by KOH etching. The remaining bulk silicon is etched with 1:3:8 HNA (hydrofluoric acid, nitric acid, acetic acid) which stops at the epitaxial layer of the device. Alternatively, a dry etching process can be used in place of the KOH to remove the 90% of bulk silicon. One such dry etch process is deep reactive ion reactive ion etching (DRIE).

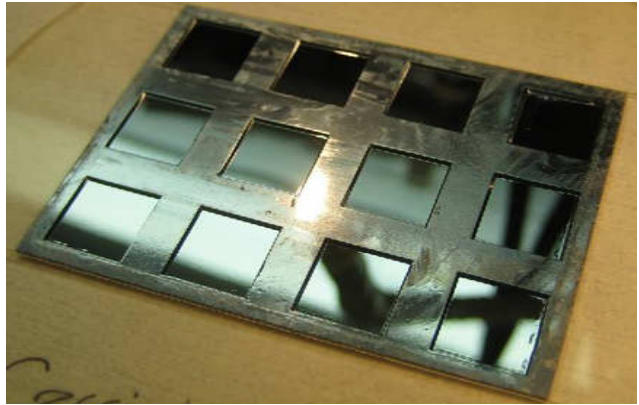

FIG. 4. A raft containing twelve thinned and delta-doped Cassini n-channel CCDs.

Cassini CCDs were used for the characterization experiments presented here. These 1Kx1K, n-channel devices were thinned to approximately 10-20μm (depending on epi thickness) and delta doped. Characterization was conducted from the VUV to the visible wavelengths.

**D. Reflectance Limit of Silicon**

Uncoated devices were used in this study to get silicon reflection-limited response. This serves a dual purpose of allowing with minimal variables the optimization of the test setup and delta doping process. Antireflective coatings can be deposited to enhance QE response, but the coated devices have more unknowns relative to bare devices. However, after adding coatings one has the added burden of decoupling the coating response from the device response, which is not trivial in the VUV. Modeling of the silicon reflection-limited response is performed with TFCALC[TM], a software package useful for modeling expected reflection, transmission, and absorption of various light wavelengths incident upon different materials.[22] Note that modeling is only as good as the database used for optical constants and most databases do not include Deep UV (DUV) optical constants.

## VIII. RESULTS

Fig. 5 shows the QE calculated from a single delta doped Cassini device. It demonstrates near silicon reflectance-limit response and so provides excellent verification of the measurement integrity. Here, quantum yield data from *Canfield, et. al.* was used. Extrinsic direct QY measurements and will be discussed in a future publication.

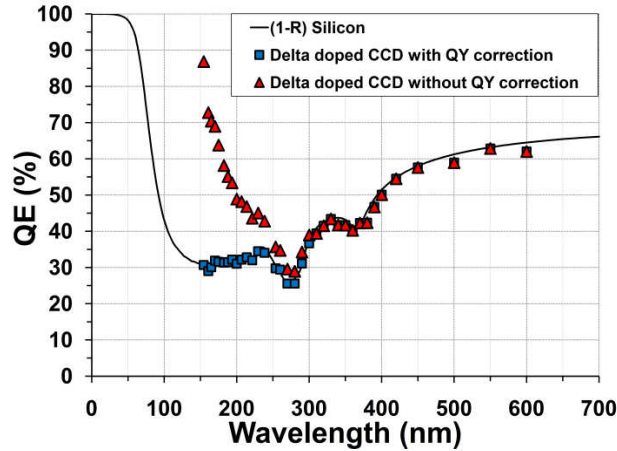

FIG. 5. Quantum efficiency measurements of thinned and delta-doped Cassini imager with modeled transmission (1-R) results. The device shows near silicon reflection limit response.

## IX.   SUMMARY

In this paper we have reviewed system requirements for QE measurements, have described JPL VUV QE characterization setup, and presented results of QE testing of delta-doped silicon CCDs at JPL. Accurate measurement of QE requires attention to the many subtleties of testing setup including source selection, filter selection, cleaning and bakeout procedures, among other issues. Critical to accurate QE measurement below 340nm is an accurate determination of quantum yield. QE measurements at JPL show excellent agreement with reflection-limited curves and high-quality from VUV through 600nm. In future publications we will examine the response of anti-reflective coatings for the 100-300nm range on delta-doped devices. Tests of this nature initially involve partial AR coating of devices so the response of bare silicon can be directly compared.

## X.   ACKNOWLEDGEMENTS

We gratefully acknowledge the Patrick Morrissey for discussions on quantum yield and Tom Elliott for support of Cassini devices. We thank Dr. Geoffrey James for information and great discussions related to the measurement system as well generous equipment contributions. We also thank the reviewer of this manuscript for extensive and thoughtful comments that improved the quality of this paper. The research was carried out at the Jet Propulsion Laboratory, California Institute of Technology, under a contract with the National Aeronautics and Space Administration.